\documentclass[manuscript, letterpaper]{aastex6}

\hypersetup{
  colorlinks = false,
}
\makeatletter 
\long\def\frontmatter@title@above{
\vspace*{-\headsep}\vspace*{\headheight}
\noindent\footnotesize
{\noindent\footnotesize\textsc{\@journalinfo}}\par
{\noindent\scriptsize Preprint typeset using \LaTeX\ style AASTeX6 v.\ 1.0 but heavily modified by David W. Hogg (NYU).
}\par\vspace*{-\baselineskip}\vspace*{0.625in}
}%
\makeatother
\sloppypar

\newcommand{\acronym}[1]{{\small{#1}}}
\newcommand{\project}[1]{\textsl{#1}}
\newcommand{\sdssiii}{\project{\acronym{SDSS-III}}}
\newcommand{\apogee}{\project{\acronym{APOGEE}}}
\newcommand{\aspcap}{\project{\acronym{ASPCAP}}}
\newcommand{\gaiaeso}{\project{Gaia--\acronym{ESO}}}
\newcommand{\galah}{\project{\acronym{GALAH}}}
\newcommand{\thecannon}{\project{The~Cannon}}
\newcommand{\foreign}[1]{\textsl{#1}}
\newcommand{\etal}{\foreign{et~al.}}
\newcommand{\documentname}{\textsl{Article}}
\newcommand{\sectionname}{Section}
\renewcommand{\figurename}{Figure} 

\newcommand{\teff}{T_{\mathrm{eff}}}
\newcommand{\logg}{\log g}

\newlength{\figwidth}
\newcommand{\insanefigure}[1]{\rule{0ex}{\headsep}\\
\includegraphics[width=\figwidth]{./figs/#1_MgFe.pdf}%
\includegraphics[width=\figwidth]{./figs/#1_MgK.pdf}%
\includegraphics[width=\figwidth]{./figs/#1_AlMg.pdf}\\
\includegraphics[width=\figwidth]{./figs/#1_NaO.pdf}%
\includegraphics[width=\figwidth]{./figs/#1_CN.pdf}%
\includegraphics[width=\figwidth]{./figs/#1_SAl.pdf}\\
\includegraphics[width=2\figwidth]{./figs/#1_gal.pdf}%
\includegraphics[width=\figwidth]{./figs/#1_HR.pdf}\\
\includegraphics[width=2\figwidth]{./figs/#1_vlon.pdf}%
\includegraphics[width=\figwidth]{./figs/#1_context.pdf}}

\newcommand{\totalnumber}{98,462}

\shorttitle{Chemical tagging can work}
\shortauthors{Hogg, Casey, Ness, \etal}
\submitted{Submitted to \textit{The Astrophysical Journal}}

\begin{document}\sloppy\sloppypar

\title{Chemical tagging can work: \\
       Identification of stellar phase-space structures \\
       purely by chemical-abundance similarity\vspace{-7ex}}
\author{David~W.~Hogg\altaffilmark{1,2,3,4},
        Andrew~R.~Casey\altaffilmark{5},
        Melissa~Ness\altaffilmark{4},
        Hans-Walter~Rix\altaffilmark{4},
        Daniel~Foreman-Mackey\altaffilmark{6,7},
        Sten~Hasselquist\altaffilmark{8},
        Anna~Y.~Q.~Ho\altaffilmark{9},
        Jon~A.~Holtzman\altaffilmark{8},
        Steven~R.~Majewski\altaffilmark{10},
        Sarah~L.~Martell\altaffilmark{11},
        Szabolcs~M\'esz\'aros\altaffilmark{12},
        David~L.~Nidever\altaffilmark{13},
        Matthew~Shetrone\altaffilmark{14}}
\altaffiltext{1}{Simons Center for Data Analysis, 160 Fifth Avenue, 7th floor, New York, NY 10010, USA}
\altaffiltext{2}{Center for Cosmology and Particle Physics, Department of Physics,
  New York University, 4 Washington Pl., room 424, New York, NY 10003, USA}
\altaffiltext{3}{Center for Data Science, New York University, 726 Broadway, 7th floor, New York, NY 10003, USA}
\altaffiltext{4}{Max-Planck-Institut f\"ur Astronomie, K\"onigstuhl 17, D-69117 Heidelberg, Germany}
\altaffiltext{5}{Institute of Astronomy, University of Cambridge, Madingley Road, Cambridge CB3~0HA, UK}
\altaffiltext{6}{Sagan Fellow}
\altaffiltext{7}{Department of Astronomy, University of Washington, Box 351580, Seattle, WA 98195, USA}
\altaffiltext{8}{New Mexico State University, Las Cruces, NM 88003, USA}
\altaffiltext{9}{Astronomy Department, California Institute of Technology, MC 249-17, 1200 East California Blvd, Pasadena, CA 91125, USA}
\altaffiltext{10}{Department of Astronomy, University of Virginia, Charlottesville, VA 22904-4325, USA}
\altaffiltext{11}{School of Physics, University of New South Wales, Sydney 2052, Australia}
\altaffiltext{12}{ELTE Gothard Astrophysical Observatory, H-9704 Szombathely, Szent Imre Herceg st. 112, Hungary}
\altaffiltext{13}{Steward Observatory, 933 North Cherry Ave, Tuscon, AZ 85719, USA}
\altaffiltext{14}{University of Texas at Austin, McDonald Observatory, USA}

\begin{abstract}
Chemical tagging promises to use detailed abundance measurements to
identify spatially separated stars that were in fact born together
(in the same molecular cloud), long ago.
This idea has not yielded much practical success, presumably because of the
noise and incompleteness in chemical-abundance measurements.
We have succeeded in substantially improving spectroscopic
measurements with \thecannon, which has now delivered $15$ individual
abundances for $\sim 10^5$ stars observed as part of the
\apogee\ spectroscopic survey, with precisions around 0.04~dex.
We test the chemical-tagging hypothesis by looking at clusters in abundance space
and confirming that they are clustered in phase space.
We identify (by the k-means algorithm) overdensities of stars
in the 15-dimensional chemical-abundance space delivered by \thecannon,
and plot the associated stars in phase space.
We use \emph{only} abundance-space information (no positional information) to identify stellar groups.
We find that clusters in abundance space are indeed clusters in phase space.
We recover some known phase-space clusters and find other interesting structures.
This is the first-ever project to identify phase-space structures
at survey-scale by blind search purely in abundance space;
it verifies the precision of the abundance measurements delivered by \thecannon;
the prospects for future data sets appear very good.
\end{abstract}


\clearpage
\section{Introduction}\label{sec:intro}

Ensembles of stars in the Milky Way are born in molecular clouds,
which are presumed to be near-homogeneous in their chemical element
composition.
However, most stars are born in unbound associations, or are born in
star clusters that disperse rapidly; they will eventually end up in
very different parts of phase space in the Galaxy.
Yet, if every star preserved its photospheric element abundances over
its lifetime (at least for most elements), then stars of common birth
origin ought to be identifiable through their detailed photospheric
abundances, long after any spatial proximity has
vanished.

This idea---dubbed ``chemical tagging''
\citep{freeman, jbh}---is one of the principal motivations for a number
of surveys, including \apogee\ \citep{apogee},
\gaiaeso\ \citep{gaiaeso} and \galah\ \citep{galah}.
In order to determine the precise abundance labels for chemical
tagging, these surveys are each measuring high-resolution, high
signal-to-noise ($S/N$) spectra for hundreds of thousands of stars across the
Galaxy's disk, bulge and halo.

Chemical tagging holds the promise of revealing not just the
star-formation history of the Galaxy, but also the accretion history
(as things that fall in are expected to be chemically distinct from
those that form in the parent body; for example, 
\citealt{eggen_1970,font,De_Silva_2007,bubar_king_2010}) and
stellar-orbit diffusion processes like radial mixing and radial
migration (for example, \citealt{roskar, quillen}).  After
stars are born---or after a star cluster is accreted and
disrupted---associations or groups will disperse, through
two-body mechanisms, interactions with resonances, or tides from the whole Galaxy.

Although undeniably promising---and motivating the 
launch of costly large-scale spectroscopic surveys---%
chemical-tagging as a search technique has yet to be proven in
practice: identifying stars of common birth origin {\it purely} on the
basis of their near-identical abundances patterns, without any
consideration of position or velocity.

Part of the reason that chemical tagging remains unrealized
is because the level of abundance specificity required
is very high.
If there are thousands of (relevant) molecular clouds forming stars in
the recent history of the Milky Way, clumps of stars can only be
identified in abundance space if abundance space is high in dimensionality.
(In principle, it needs to be high in dimensionality both in terms
of the number of measured abundances \emph{and} in terms of the number of
nucleosynthetic pathways or the dimensionality of the true abundance
space.)
Therefore accurate---or at
least precise---measurements of many different abundances are needed
for stellar siblings to have sufficiently unique fingerprints.
Stellar spectroscopic surveys now have the resolution,
$S/N$, wavelength coverage, and sample sizes to deliver many
different chemical tags for each star.

There are, however, two big issues.
The first is that the physical assumptions behind the idea may require
refinement:
There may be chemical-abundance overlaps among open clusters
\citep{blancocuaresma}, coeval groups of stars may have similar tags
but different birth places \citep{mitschang}, and the
chemical-abundance space might be low in dimensionality.
On the other hand, precise studies of stellar twins \citep{melendez, jofre}
indicate that pairs of stars can be found with unusually similar abundances, 
open clusters show remarkably uniform chemical abundances \citep{bovy},
and peculiar abundance ratios
have been successfully used to identify disrupted cluster members (for example, \citealt{ocen}).
There have also been hints seen of relationships between chemistry and kinematics
(for example, \citealt{helmi, helmi2}).

The second issue for chemical tagging---and the one we address
here---is measurement precision (and accuracy).
The current precision on abundance measurements in the published
survey catalogs is not high enough (for example, \citealt{martell, ting}).
However, our recent work \citep{thecannon} suggests that the \emph{data}
are precise enough: There is enough $S/N$ at the relevant
locations in spectrum space to deliver high-precision tags.
The existence of very large data sets, homogeneous in spectrum
space, suggests that data-driven approaches to the determination
of stellar abundances might considerably outperform traditional
methods.
These traditional approaches are based on \foreign{ab-initio} physical models
that have shortcomings that become apparent in this age of
high-quality spectra, and the data-analysis methods do not make use of
all of the information in the data sets.
Improving the models and exploiting the entire information
content in the data is critical if we are going to deliver useful
chemical tags.

Our specific contribution in this space has been to develop
\thecannon\ \citep{thecannon, ages}, which is a data-driven model for
stellar spectra.
This model can deliver stellar parameters and chemical abundances for
stars, making use of every pixel of every stellar spectrum (that is,
all the information in the data) but making no use at all of physical
models of stars.
It relies only on there being training data---\emph{some} reference stars for
which parameters and abundances are known and believed.
In companion papers \citep{casey16, ness16} we show that
\thecannon\ can deliver 15 to 19 abundances for stars in the
\apogee\ survey, at precisions higher than even the training (reference) data on
which the model is trained.
We say more about this below.
We will show here that \thecannon\ improves chemical abundance measurements
to the point that \emph{chemical tagging is now possible.}

One note on \emph{accuracy} and \emph{precision:}
In principle, the problem of chemical tagging does not require
absolute accuracy for chemical-abundance measurements.
It only requires that we can precisely see that two stars are similar
in their abundances, even if we happen to be wrong about the absolute values of
those abundances.
This point might make it seem like we don't care that our models are
wrong, so long as they are \emph{consistent.}
However, this is a bit misleading:
For chemical tagging to succeed, we need stars with different
atmospheric parameters $(\teff, \logg)$ but the same chemical
abundances to be assigned the same position in chemical-abundance
space.
That doesn't require \emph{overall} accuracy, but it requires that the models
have the right dependencies on atmospheric parameters such that the
wrongness in abundance space is consistent across the Hertzsprung-Russell diagram.
That is, we need a substantial amount of accuracy to achieve our goals.

It is important to realize that analysis techniques based on \foreign{ab-initio}
physical models are inaccurate, yet they too strive to improve precision
in the face of knowingly inaccurate models.  Incomplete atomic data, 
simplifications of photospheric structure, assumptions about convective
motion, and inconsistencies resulting from positing local thermal
equilibrium all contribute to produce inaccurate abundances.
\thecannon\ stands out because it demonstrably improves the precision on
chemical abundances, whereas the accuracy of those labels is limited only
by the training (reference) set employed.  At the same time, it is crucial
to be cognisant of the constraints in the training set: the
results of \thecannon\ will be limited by the quality of the training 
set labels. While chemical tagging does not necessarily require
accurate abundance labels (precision is paramount), comparing abundance
labels to models of Galactic chemical enrichment requires a firm level
of belief in the label accuracy. Particularly for the abundance labels
of stars in the training set.
    
In what follows, we are going to use \apogee\ \acronym{DR12}
data \citep{apogeedr12, dr12}
from \sdssiii\ \citep{sdssiii}, in
which we can re-derive 15 element abundances (C, N, O, Na, Mg, Al, Si,
S, K, Ca, Ti, V, Mn, Fe, Ni), using \thecannon\ \citep{thecannon}.
The \apogee\ data set covers a huge radial extent and---because the
data are taken in the infrared---is capable of exploring all parts of the disk,
including the thin parts.
However, it has the disadvantage that its spatial coverage is incomplete
(that is, limited pointings produce a fractured sky map),
which makes it hard to see, within the data set, linear or
extended stellar structures.
In many ways, \galah\ will deliver improvements, both because it will
have more abundances (possibly 29) and contiguous sky coverage \citep{galah}; that
said, it will not observe much of the thin disk.

One final note:
we think of this \documentname\ as performing a proof of concept.
We know that the stellar members of open and globular clusters---stars
that are identified by being close in phase space---contain highly
informative abundance information that identifies them also in
chemical-abundance space.  Does this work the other way around? 
Can chemical tagging identify small subsets of stars, among a vastly
greater background sample, that have a common birth origin? If we 
find stars purely by their clustering in abundance space and subsequently
show that they are part of a still spatially coherent cluster, group, or stream, then
we will have resolved all practical outstanding issues plaguing 
chemical tagging, thereby bringing us much closer to
unravelling the formation of the Milky Way.

\section{Data: \apogee\ giants with abundances from \thecannon}\label{sec:data}

Our analysis draws on the spectra of \totalnumber\ giant stars ($\logg
< 3.9$) from \apogee\ \acronym{DR12} \citep{apogeedr12}, with no warning flags
set in the \apogee\ \aspcap\ \citep{aspcap} pipeline
reductions.
We re-analyze these \emph{spectra} using \thecannon, because it can
deliver stellar abundance labels of higher precision, especially for
stars of $S/N \le 150$.
The details about how we select, reduce, and analyze the \apogee\ data
are given in full detail in the companion papers \citep{casey16,
  ness16}, and we only summarize briefly here:
We re-derived 17 labels ($\teff$, $\logg$, and 15 abundances
referenced to H) from the \apogee\ \acronym{DR12} spectra. 
For the training step of \thecannon, 
12,681 red-giant stars with spectral $S/N \ge 200$ were used.
For this small fraction of the sample with the highest $S/N$, the
labels provided by \aspcap\ provide consistent, relatively low-scatter, sensible
abundance-space measurements.
The training step fixes a spectral model
that predicts the normalized spectrum as a quadratic function of all
labels ($\sim 1.5\times10^6$ model coefficients). In the second stage---the test step---each unlabelled star is
used to establish a single-star likelihood function for its labels,
holding the spectral model coefficients (the parameters of the model) fixed.
\thecannon\ finds the labels that minimize the single-star likelihood
function for each test-set star.
This optimization is not convex, but it is trivially parallel,
and therefore fast. The test
step for 150,000 spectra takes less than 30~minutes on a small 
research cluster in Cambridge.

It is important here to note some of the limitations of stellar labels
delivered by \thecannon.
The first is that \thecannon\ is only as good as its training set!
All of the biases and calibration issues in the input training set will be
delivered to the output labels.
This means that there is no sense in which \thecannon\ delivers
absolute abundances any better than \aspcap\ does.
The second is that \thecannon\ (in the form used here) is aggressively
data-driven.
It will use anything it can to measure, say, the Na abundance, not just
actual Na lines.
This means that population-level correlations in  element abundances are
being used to deliver information about individual elements.
We bring up Na as an example here for the important reason that at low
metallicity, there are no significant Na lines in the spectra; we are
measuring Na only indirectly at low metallicity.
A third limitation of \thecannon\ is that it is operating a regression
in a very high dimensional label space.
This regression is hard to test and validate near the edges of the range
of applicability.
One way to say this is that the convex hull of the training set of
points in 17-dimensional space is potentially very small, and it is
hard to visualize what is going on outside that hull.
For these reasons, there might be label biases that grow with displacement
from the bulk of the training data.

\begin{figure}[!p]
\insanefigure{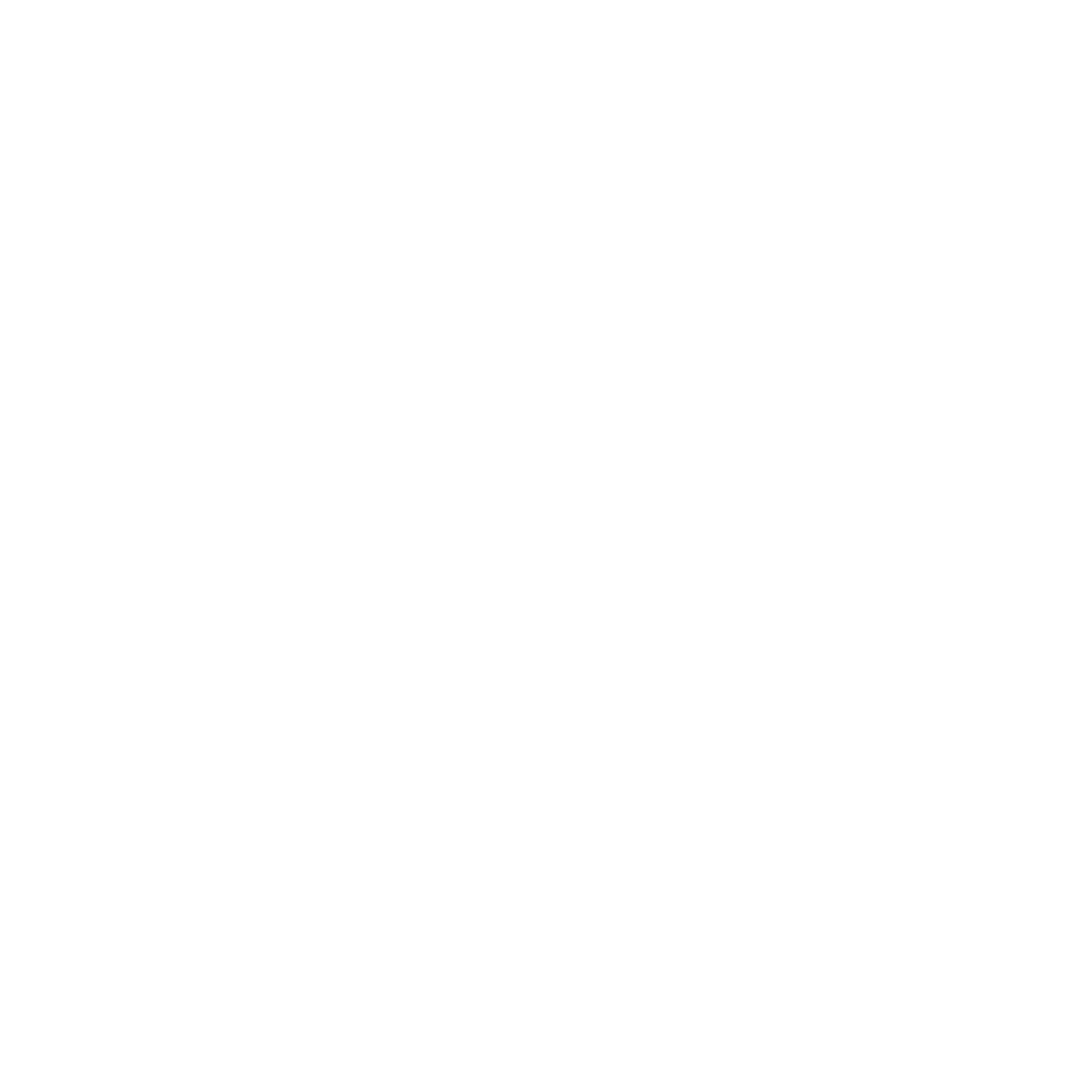}
\caption{The full sample of \totalnumber\ stars used in this study.
  The top six panels show six two-dimensional projections of the
  empirical abundance space distribution.
  The clustering algorithm (described in
  \sectionname~\ref{sec:method}) performs clustering in a full
  15-dimensional abundance space of elements referenced to H (not Fe),
  but plots are shown here referenced to Fe for familiarity reasons.
  The bottom three panels show stellar meta-data not used in the
  clustering described below.\label{fig:all}}
\end{figure}

Six two-dimensional projections of the abundance data, plus some other
data quantities, are shown in \figurename~\ref{fig:all}.  Throughout
this work we show two-dimensional abundance projections for only a few
elements. The entire sequence of 15-choose-2 combinations is too
immense to visualize. The projections shown are ones which
have been extensively used in abundance studies of globular clusters
and satellite systems. These are typically light element abundances,
and our projections include C--N, Na--O, and Mg--Al. Globular clusters
famously demonstrate correlations in these elements \citep[for example,][and references therein]{Norris_1995,Carretta_2009}, thereby making
them suitable projections for us to show in verifying and examining
any substructure identified by k-means (or any other clustering algorithm).

\clearpage
\section{Identifying abundance-space clusters}\label{sec:method}

Although only hints of small-scale clustering in the abundance space
are visible in \figurename~\ref{fig:all}, exploration of the data by hand indicates that
known clusters do appear in the high-dimensional abundance space as over-densities.
In general, the collapse of 15 dimensions down to 2-d projections
will hide, smooth, or dilute any structure; there is no guarantee that
a highly featured distribution in 15-d will show features in
\emph{any} 2-d projection, let alone axis-aligned, human-selected projections.
This encourages us to look for over-densities automatically in the
abundance space and see if anything found that way would be over-dense
in phase space.
The simplest method for clustering points in $D$-dimensional space is
the \emph{k-means} algorithm (see \citealt{bishop} for a pedagogical
introduction and references to the original literature).

Briefly, the k-means algorithm is the following:
$\bullet$~Start at some initial guess for the locations of $K$ $D$-space cluster centers ($K$
locations in the $D$-dimensional space).
$\bullet$~Assign each point in the
space (each star in our case) to the closest of the $K$ centers.
$\bullet$~Given this assignment of stars to centers, update each center (each of
the $K$ $D$-space positions) by taking the mean of the locations of
the stars assigned to that center.
$\bullet$~Iterate these steps (assignment of points, followed by taking of means) to convergence.
The output of this algorithm is the converged locations of the $K$ centers and the assignments of
all points to those centers.
This algorithm is fast and performs well in practice in problems of
this nature; also we are not the first to use the k-means algorithm in
abundance space \citep{gratton}.

The k-means algorithm has a number of limitations:
One is that $K$ must be chosen by hand (or heuristically at best).
Here we are only demonstrating a concept; we don't need to have the
best possible clustering.
For this reason we simply choose $K=128$, $K=256$, and $K=512$ and look at all the
results.
Also, the k-means algorithm only performs local optimization.
At each $K$ we perform 32 restarts with different initializations, and
preserve the best clustering (best according to the k-means score).
Each of the 32 initializations was performed with the
\project{scikit-learn} standard k-means initialization procedure
\citep{sklearn}.

Another issue with k-means is that it effectively uses metric
distances in the $D$-space; it presumes Euclidean isotropy.
We choose here to work in the hydrogen-normalized abundance space, the
space of [C/H], [N/H], [O/H], [Na/H], [Mg/H], [Al/H], [Si/H], [S/H],
[K/H], [Ca/H], [Ti/H], [V/H], [Mn/H], [Fe/H], [Ni/H].
But in addition to this, we re-scale these by approximate typical measurement
precisions obtained by \thecannon\ before running k-means.
These scalings were
  [C/H]/0.041,
  [N/H]/0.044,
  [O/H]/0.037,
  [Na/H]/0.111,
  [Mg/H]/0.032,
  [Al/H]/0.055,
  [Si/H]/0.041,
  [S/H]/0.054,
  [K/H]/0.069,
  [Ca/H]/0.043,
  [Ti/H]/0.072,
  [V/H]/0.146,
  [Mn/H]/0.041,
  [Fe/H]/0.019, and
  [Ni/H]/0.034.
This scaling makes the space close to isotropic in measurement uncertainty or
observational precision.
Finally and related, because it looks for clusters compact in
metric distance, k-means is more sensitive to clusters that are spherical
in the scaled abundance space than clusters of the same density
that are elongated in any sense.
Importantly we use \emph{only} abundance-space information, and no
positional or velocity information (nor $\teff$ nor $\logg$ nor any
targeting or observational meta-data) as input to the clustering
algorithm.

Nothing about this algorithm or our choices are particularly
tuned or optimized; this is in no sense the
algorithm or the method that reveals the best structures.
We chose k-means as a simple and straightforward approach to identifying
clusters in high dimensional label space, with very few control parameters
or decisions required.
It is also an algorithm that is well studied in the machine-learning literature,
so it has well understood properties.
In practice more complex
clustering algorithms may perform better than k-means, or even
tuning k-means by heuristically setting $K$ is likely to improve
upon the results here. This \documentname\ serves as a proof of concept.
Indeed, it is a feature of this work that even the simplest, most
generic clustering algorithm returns interesting
structures (as shown below). 
The prospects for Milky Way science only improve as the clustering
algorithm improves.

\begin{figure}[!bp]
\includegraphics[width=1.5\figwidth]{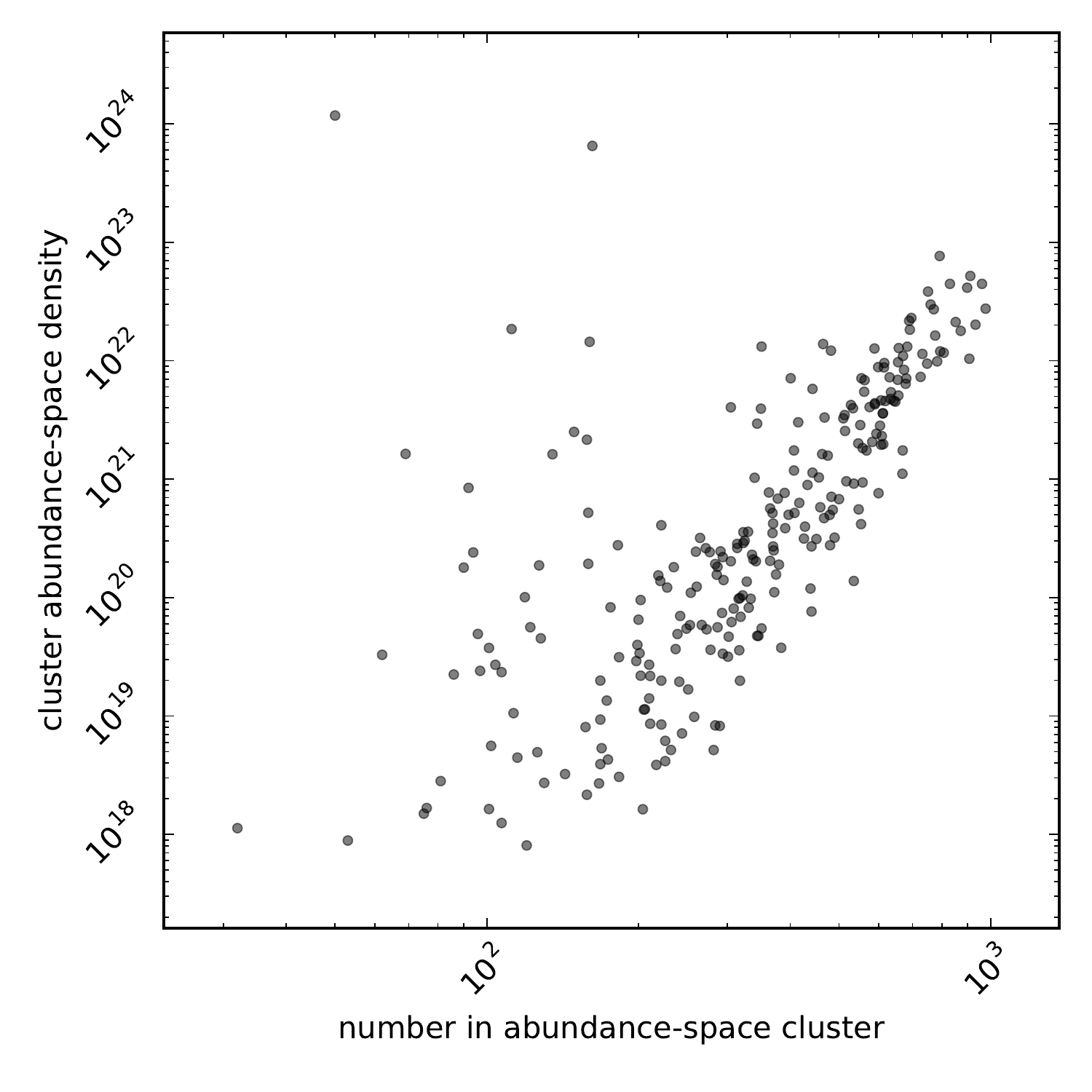}
\caption{The distribution of membership and density for the 256
  abundance-space clusters returned by the k-means algorithm at
  $K=256$.  There is a bulk trend and then clusters that are much more
  dense than the trend. The densest cluster is displayed in more detail in 
  \figurename~\ref{fig:M13}.\label{fig:densities}}
\end{figure}

When the k-means results are returned, the $D\times D$ empirical
variance tensor for the members of each cluster can be constructed.
From this, an effective density in the abundance space can be computed
as the number of points in the cluster divided by the square-root of
the determinant of the tensor.
This density was used to rank abundance-space overdensities
for visual inspection.
In \figurename~\ref{fig:densities} we show the distribution of cluster membership
and density for the $K=256$ run of k-means.
There is a bulk trend of larger clusters (clusters with more members or higher occupation number)
being more dense, but the most interesting
k-means densities are those that are more dense than this trend.

Importantly, the k-means algorithm assigns \emph{every} star to at
least one cluster.
For this reason, there is no sense in which \emph{every} ``cluster'' returned
by k-means is a distinct overdensity in abundance space.
In what follows, we only consider high-density clusters---clusters
that are more dense than average for their occupation number (total membership); these ought to represent true
over-densities in abundance space.

\begin{figure}[!p]
\insanefigure{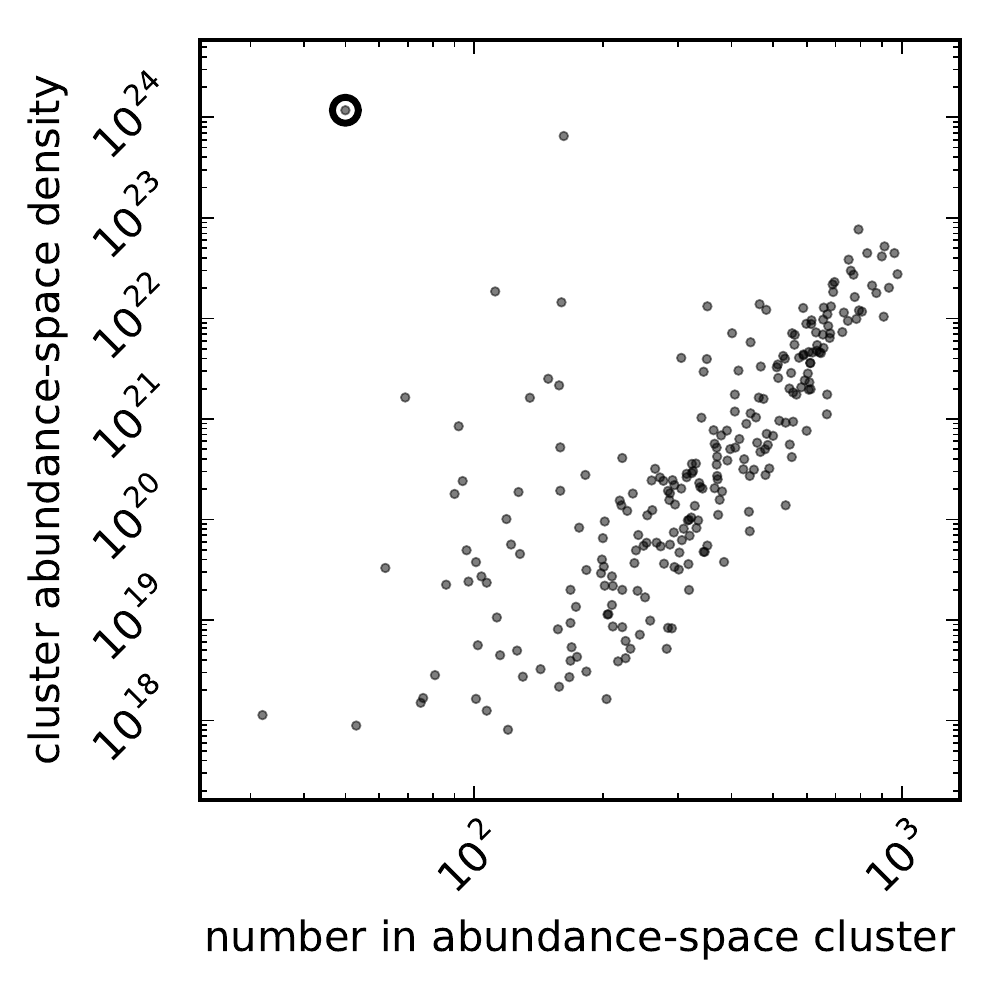}
\caption{Same as \figurename~\ref{fig:all} except that the entire
  sample has been made gray and the members of the most dense $K=256$
  abundance-space cluster have been rendered as unique, prominent
  triangles (color is velocity rank, orientation is $\logg$ rank).
  The lower-right plot shows this cluster (circle) in context of the
  other clusters (dots).  This cluster, which was identified only in
  abundance space (six projections of which are the top six panels in
  this \figurename), turns out to be dominated by the halo globular
  cluster M13. The symbol orientations and colors have no meaning;
  they are randomly generated---one unique color and orientation for
  each highlighted star---to make the points cross-identifiable across
  panels.\label{fig:M13}}
\end{figure}

\begin{figure}[!p]
\insanefigure{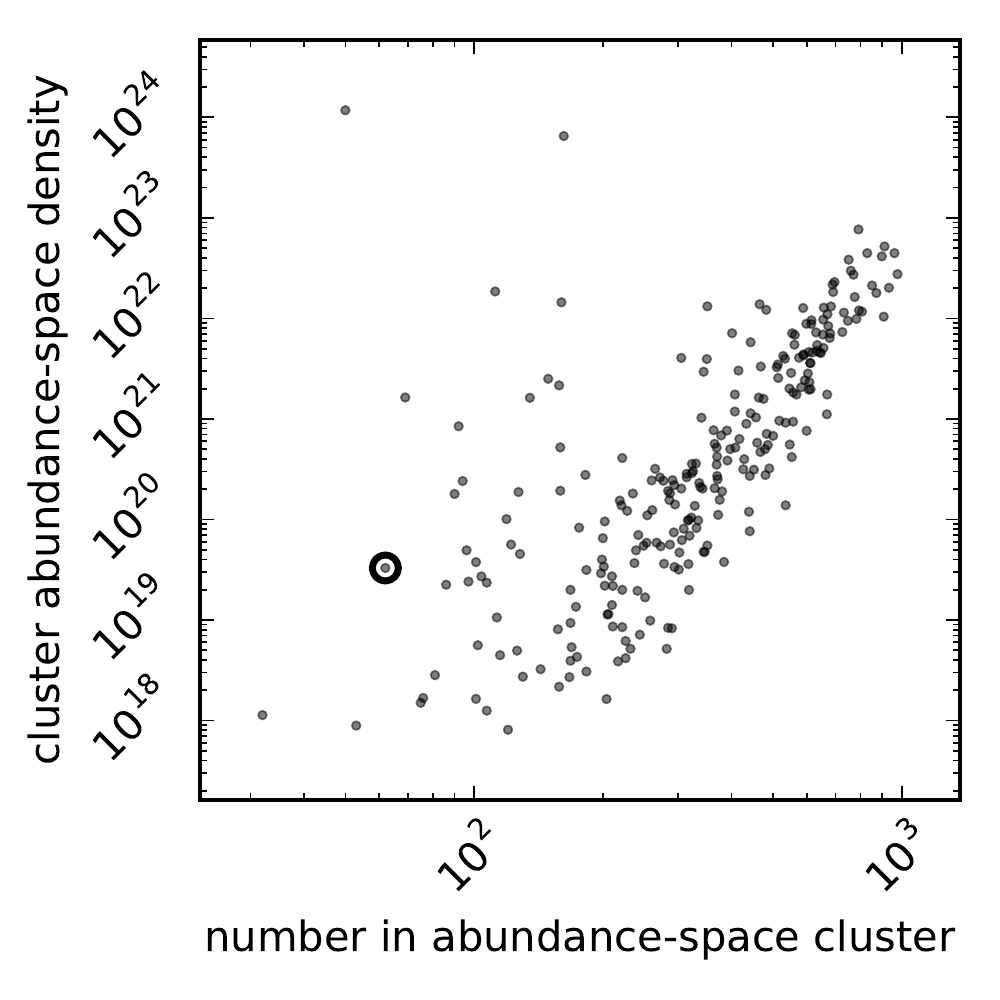}
\caption{Same as \figurename~\ref{fig:M13} but for another dense
  abundance-space cluster.
  This one turns out to be dominated by globular
  cluster M5.\label{fig:M5}}
\end{figure}

\begin{figure}[!p]
\insanefigure{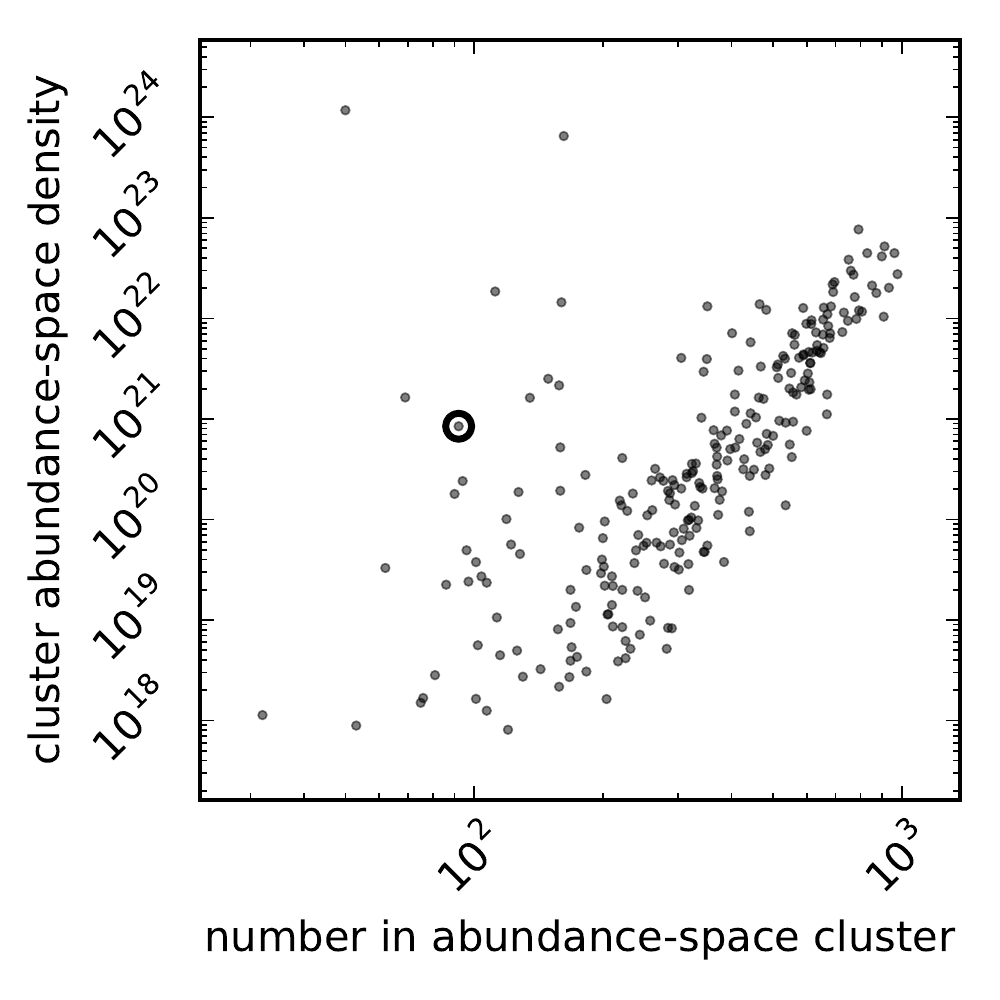}
\caption{Same as \figurename~\ref{fig:M13} but for another dense
  abundance-space cluster.
  This one turns out to be dominated by the Sagittarius dwarf spheroidal galaxy.\label{fig:Sgr}}
\end{figure}

\begin{figure}[!p]
\insanefigure{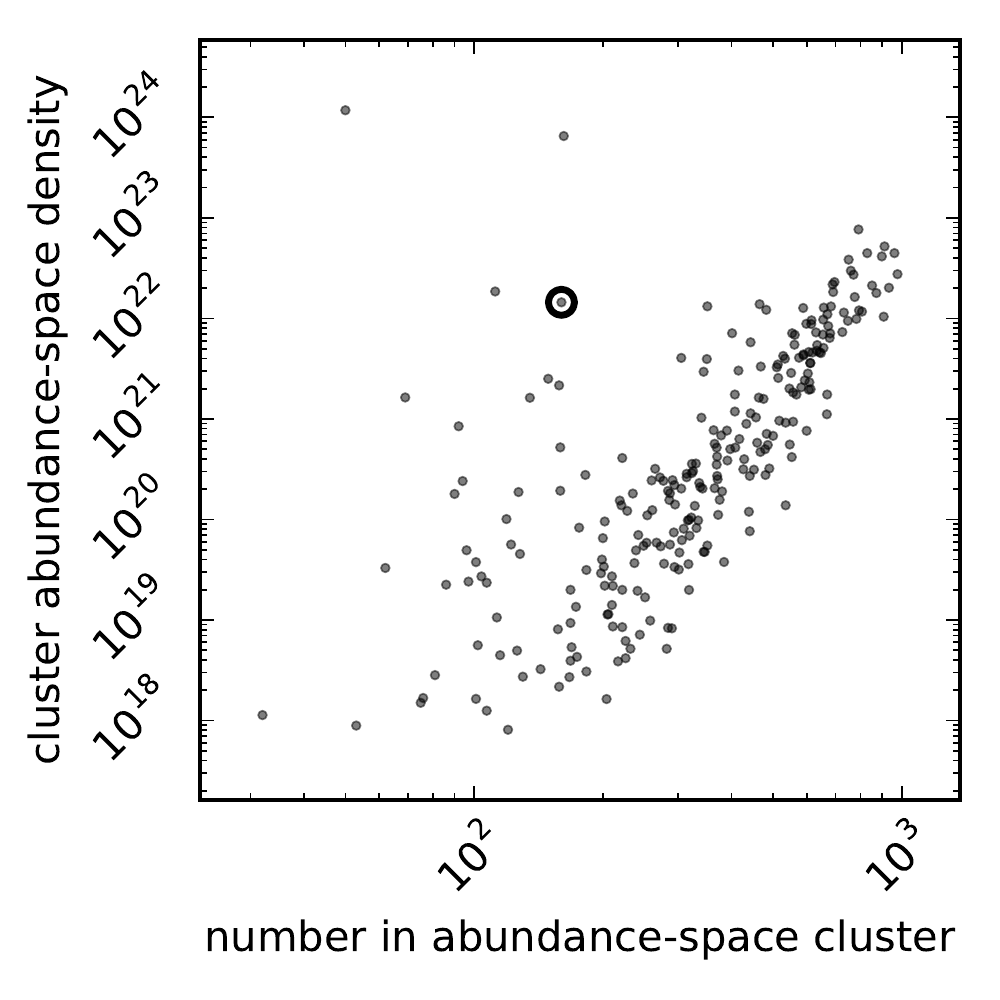}
\caption{Same as \figurename~\ref{fig:M13} but for another dense
  abundance-space cluster.
  This one turns out to be dominated by a hitherto unrecognized high-velocity-dispersion
  structure in the Galaxy halo.\label{fig:halo}}
\end{figure}

\begin{figure}[!p]
\insanefigure{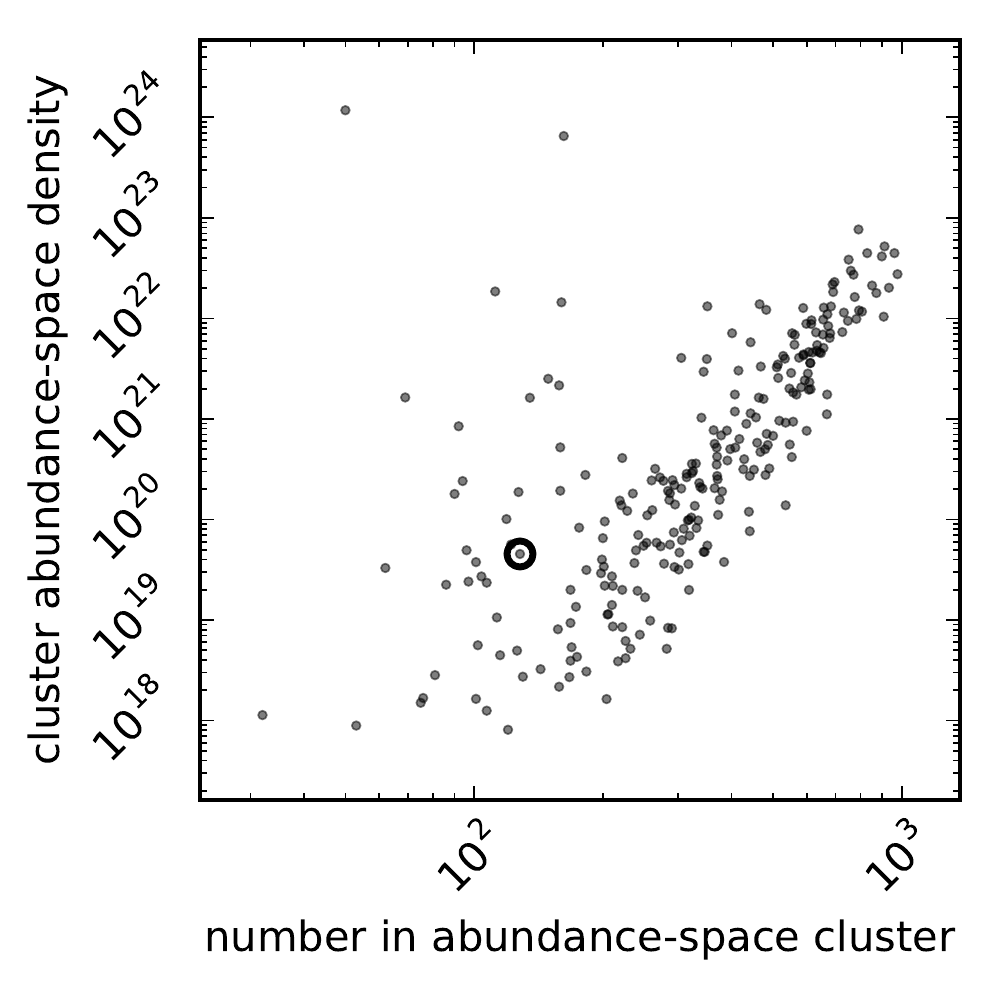}
\caption{Same as \figurename~\ref{fig:M13} but for another dense
  abundance-space cluster.
  This one turns out to be dominated by a very thin young stellar
  structure in the Galaxy disk, probably one that has been discovered previously \citep{wegg}.\label{fig:disk}}
\end{figure}

We chose a few interesting cases from the high-density clusters
in the $K=256$ run and show them in
\figurename~\ref{fig:M13}, \ref{fig:M5}, \ref{fig:Sgr},
\ref{fig:halo}, and \ref{fig:disk}.
The first three of these are dominated by stars in the known clusters
M13 and M5, and the Sagittarius stream (we identify overlap with these objects
by looking at stellar position and velocity, and, in some cases,
\apogee\ targeting flags).
The fourth is a halo structure with high velocity dispersion and
possibly accreted.
The fifth is a thin-disk star-formation feature.
We will discuss the astrophysical implications of these results in
\sectionname~\ref{sec:discussion}.

The abundance-space clusters shown in the \figurename s
might not look dense in 2-d projections of the abundance space, but they
are very dense in the 15-dimensional space.  Indeed,
\figurename~\ref{fig:M13} is the densest cluster found in the 15-d
space \emph{by far}.
The challenge of this work is to find structure in the high
dimensionality space that is not obviously visible in any 2-d projection.
Although we have by no means any model of the 15-d space, we do know
that the structures shown in the \figurename s are high in 15-d
density, or at least relative to other clusters with the same occupation number.

We have only shown results from the $K=256$ run of k-means.
This run was chosen because its densest abundance-space clusters map
well onto known stellar clusters.
At $K=128$ the densest abundance-space clusters tended to combine
multiple known stellar clusters into single, large abundance-space
groupings.
At $K=512$ k-means tended to split even mono-abundance groups into
smaller sub-groups.
This is a reminder that k-means has been chosen just for simplicity
here; it is by no means matched to the discovery of
stellar structures.
An important follow-up project is to build a model of abundance space
that captures the features expected from stellar populations (and observational uncertainties).
Even if we were able to tune k-means in some sense ``perfectly'', it
would still break up many stellar clusters, since they often show
multiple stellar populations and non-linear correlations between element abundances.
It is also the case that each of the abundance-space clusters we
\emph{do} show in the \figurename s includes both star-cluster members
and some background contamination, and also is missing some true
star-cluster members.

Although we show only three stellar clusters in
\figurename~\ref{fig:M13}, \ref{fig:M5}, and \ref{fig:Sgr}, many
other clusters are visible among our densest abundance-space k-means
clusters.
These include M15, M92, and M107, among others.
We haven't asked yet whether we detect---as abundance-space
overdensities---all of the stellar clusters we expect to find. (Many
clusters within the \apogee\ data set only contain a few plausible 
members, thereby  complicating any inferences we wish to make about 
detection completeness using simple k-means.) 
This \documentname\ is simply a demonstration of the chemical-tagging
concept; a full investigation of whether we can construct complete
catalogs of stellar clusters is beyond our present scope.


\clearpage
\section{Discussion}\label{sec:discussion}

We have demonstrated for the first time that, in a large-scale
spectroscopic survey, stars that appear similar in
chemical-abundance properties are often associated physically in
clusters or other structures.
The reverse is well established---that is, that stars that are
associated physically in clusters are similar in chemical-abundance
properties.
That is, although they often show chemical diversity, that chemical
diversity is small, and follows well-defined trends or a subspace in
abundance space (for example, \citealt{gratton, meszaros, bovy},
and references within those).
However, it was not known---prior to this work---whether abundance-space information would be
informative enough, or \emph{measurable} precisely enough to find this
structure in surveys dominated by a large mix of stars from different
origins and ages.
There was good reason to be pessimistic, but plausibly the increased
precision obtained here by \thecannon\ made this possible; even tiny improvements
in individual abundances can enormously increase contrast in the high-dimensional
abundance space \citep{ting}.
We have demonstrated with a straightforward (and in no way optimal)
clustering in chemical-abundance space that it is easy to find co-eval
stellar structures.
Although this is very strong evidence that chemical tagging will work,
we have not performed, in any sense, the canonical procedure of
looking for features that are consistent with being delta functions in
chemical-abundance space (convolved with a measurement noise
distribution); we have looked only for over-densities, not delta
functions.

One clear result here is that it is easier for the clustering algorithm
to find structures at low metallicity than at high metallicity:
Some of our low-metallicity features are very compact in phase space,
while none of our higher metallicity features are.
This probably relates to the much higher background
(in abundance space) of unclustered stars at higher metallicities than
at low metallicities; in general it is easier to find abundance-space features
at the edges of the distribution than in the center \citep{ting}.

What made our success here possible is substantial improvements in precision
delivered by \thecannon.
We don't have an absolute measure of the precision of the measurements,
but from looking at open and globular clusters, it appears to be on the
order of or better than 0.04~dex for the median element in the list of 15.
These improvements are discussed in detail in the companion papers
\citep{casey16, ness16} about the abundance measurements.
We believe these improvements come from a combination of factors, not
limited to (a)~improvements in the determination of pseudo-continuum,
(b)~the use of more spectral range than just unblended element windows
\citep{aspcap}, and (c)~accurate spectral predictions
(\thecannon\ delivers accurate predictions because it is fit to
observed data). 
Furthermore, because abundance space is high in dimensionality, even
small improvements in abundance measurement get taken to a significant
power when thinking about information gains.

There are a few points to highlight about the abundance anti-correlations
found in globular clusters. First, the detection of these signatures gives
confirmation to our analysis: The training (reference) set was not purposefully selected
to contain members of any globular or open cluster. However amongst the test
set, we recover globular cluster members and their anti-correlations 
in chemical abundances. That is to
say we recover peculiar chemical abundances that are not dominant in our
training set, but are expected from previous studies. Secondly, the very
presence of complex abundance correlations makes the detection of clusters
(by k-means) substantially more difficult. Briefly, the k-means algorithm
is most effective for near-normally distributed, isotropic clusters in high-dimensional space.
While globular clusters often show distributions of this kind in a few dimensions,
they also demonstrate non-linear correlations which would be
sub-optimally selected by k-means. For these reasons the fact that (a)~\thecannon\
reports these abundance correlations and (b)~k-means (as a simple algorithm
without information about expectations in abundance space) does not appear to be
severely impacted by these correlations, gives us great hope for more sophisticated approaches.
That said, globular cluster members are over-represented in the \apogee\ data set,
because many of them were specifically targeted for observations for being cluster
members \citep{apogee}; that is, globular-cluster structure is over-represented relative to the
field in this data set.

The abundances of known clusters that we identify by k-means are in
excellent agreement with the literature. For the group shown in \figurename~\ref{fig:M13},
which we attribute to M13, we find a mean of [Fe/H]$ = -1.55$.
The mean and spread agree well with other work on this cluster
\citep{Kraft_1992,Cohen_Melendez_2005,Johnson_Pilachowski_2012}, and compiled catalogs of
globular cluster properties \citep[][accessed 2016]{Harris_1996}. The detailed abundances are also
consistent: we find a correlation in [C/Fe] and [N/Fe] abundances \citep{Smith_2005}, and their
projection in \figurename~\ref{fig:M13} show a spread that reflects deep mixing
along the red giant branch \citep{Briley_2004}. We find that the light element
abundances (most notably Mg--Al) for this cluster are anti-correlated, as
expected from other studies \citep[for example,][]{gratton}. However we find a
smaller spread in [O/Fe] than reported by others who have looked exclusively
at M13 \citep{Johnson_Pilachowski_2012}.  As we discuss above, this is expected:
k-means is by no means optimal for identifying arbitrary-shaped structures
in high-dimensional space, and the first stars that would be assigned to another k-means
cluster would be those with the most extreme abundances: low [O/Fe] and
high [Na/Fe].

We find a mean metallicity of [Fe/H]$ = -1.3$ for the cluster we
associate as M~5 (\figurename~\ref{fig:M5}), which agrees well with
the [Fe/H]$ = -1.33 \pm 0.03$ measurement 
\citep{Koch_2010}, and the $-1.29 \pm 0.02$ value listed in the
standard compilation (\citealt{Harris_1996}; accessed 2016).  The correlations
in light elements---specifically Mg--Al and C--N---also agree well
with other studies \citep{Ivans_2001,meszaros}.  In particular we find
only a weak correlation in the [Na/Fe]--[O/Fe] abundance ratios
\citep{Lai_2011}.  However, when we consider the extent of the
literature on M~5, it would suggest that we do not recover the full
extent of these correlations: the stars with the highest [Na/Fe],
[Mg/Al], and lowest [O/Fe] abundance ratios are not represented in the
cluster that we associate as M~5.  This is likely a consequence of our
(poor) choice of the k-means algorithm, which is most effective for
near-circular distributions in the scaled abundance space, and reduced in power for clusters
 that have polynomial relationships in dimensional
space (for example, [Na/Fe]--[O/Fe]).  Indeed, 
abundances for 122 members of M~5 have been reported in \apogee\ \acronym{DR12} data (\citealt{meszaros}),
whereas the cluster we associate as M~5 only contains $\approx60$
members, and only covers about half of the extent of [Na/Fe]--[O/Fe]
relationship.

All that said, we reiterate a caveat here that we also mentioned in
\sectionname~\ref{sec:data}, which is that at very low metallicity, Na
does not show strong lines in any \apogee\ spectrum.  For this reason,
the [Na/Fe] shown for the low-metallicity structures are obtained not
by measuring Na lines but rather the lines of elements that correlate
strongly with Na at the population level.
The true [Na/Fe] may show correlations, variations, or anomalies in
the clusters that are not captured by \thecannon\ working at low
metallicity.

The high velocity-dispersion structure seen in \figurename~\ref{fig:halo} stands
in stark contrast to the rest of the clusters we have identified. In
addition to being highly clustered in 15-dimensions, it is visibly
clustered in our 2-d projections. Indeed, it is more clustered in our
abundance projections than the young stellar structure in the disk,
and comparably so to the other known globular clusters that we have
shown here. The disparity between this structure and others presented
here is the lack of co-spatial stars. Twelve stars within this cluster are
marked as `M53' candidates by \apogee, but the remaining 145 stars
are spread amongst fields throughout the halo.

It is conceivable that this structure, being near the outskirts of the
abundance distribution, is somehow concentrated in abundance space by
some kind of shrinkage (in the statistical sense) induced by
\thecannon's regression.
However, it is a dense feature in abundance space, and is worth
following up.
If follow-up observations show that the stars have abundances and that
are consistent with having origin in a single stellar population or
low-metallicity dwarf galaxy, it could represent an accretion event in
the Milky Way halo.

The k-means algorithm employed here is by no means optimal, and other clusters
we identify are accompanied by a few stars that are not currently co-spatial.
Those stars may be unassociated interlopers that have been misclassified,
or they may be true cluster members that are now unbound. However in the
case of the high-velocity-dispersion structure in \figurename~\ref{fig:halo},
the situation is far more extreme. We identify a group of stars with
very similar abundances in 15 dimensions that are now spread throughout
the Galaxy halo.
Because some stars ($<8$\,percent) are candidate members of the massive globular
cluster M53, it provides the tantalizing possibility that these stars may
indeed be all members of the same co-natal gas cloud, implying that 
these stars have been accreted onto the halo from a single globular
cluster early in the Milky Way's formation (as
has been hypothesized elsewhere; for example in \citealt{grebel}). While speculative,
this idea demonstrates the promise of chemical tagging. 

The abundance-space group of stars in the disk structure shown in
\figurename~\ref{fig:disk} may be associated with a very thin
Milky-Way bar component reported previously \citep{wegg}, which is
found to exist predominantly toward the end of the bar (at $l$
$\approx 30$~deg).
The stars in this structure are metal-rich and have a low
[C/N] ratio and are therefore likely young (following the [C/N]--mass
relationship; \citealt{martig}).
It is possible that this structure has very low scale height because
it is very young and hasn't experienced dynamical heating or radial
migration.

It is worth reiterating here what is written above:
There is no sense at all in which \figurename~\ref{fig:M13},
\ref{fig:M5}, \ref{fig:Sgr}, \ref{fig:halo}, and \ref{fig:disk} show
a representative or complete set of features found in abundance space.
These features were hand-chosen to be obviously interesting and
interpetable.
There are many other things to be found in this data set, and many
more effective models to be built of abundance-space structure.
The only strong conclusion of this investigation is that chemical
tagging \emph{can} work; the abundance measurements with \thecannon\ are
precise enough and the chemical-abundance space is informative enough.

The \apogee\ project shows great promise for these studies, and with
the appearance of the companion papers, we will release the
chemical-abundance measurements we used here for further study.
The future is even brighter, however: \apogee\ is expanding to more measured
elements, more stars, and all-sky angular coverage with \apogee2, and
\galah\ is working towards releasing chemical abundances on a larger set
of 29 elements.
The \apogee\ elements employed in this project include alpha, light proton-capture, odd-Z, and iron peak elements.
\galah\ will deliver element abundances from all the major
nucleosynthetic processes, including light proton capture elements,
alpha, odd-Z, iron-peak, as well as neutron-capture elements.
Chemical tagging capabilities are expected to grow as the
\emph{product} of the measured nucleosynthetic pathways.

\acknowledgements
It is a pleasure to thank an anonymous referee for useful comments that have led to improvements in the manuscript.
We also thank
  Joss Bland-Hawthorn (Sydney),
  Jo Bovy (Toronto),
  Charlie Conroy (Harvard),
  Katia Cunha (\acronym{NOAO}),
  Amina Helmi (Kapteyn),
  Jeremy Magland (\acronym{SCDA}),
  Don Schneider (\acronym{PSU}),
  Keivan Stassun (Vanderbilt),
  Angus Williams (Cambridge),
  and the Blanton--Hogg group meeting
for valuable discussions and comments.
This project was funded in part by
  the \acronym{NSF} (grants \acronym{IIS-1124794}, \acronym{AST-1312863}, \acronym{AST-1517237}),
  \acronym{NASA} (grant \acronym{NNX12AI50G}),
  the Moore-Sloan Data Science Environment at \acronym{NYU},
  the Australian Research Council (DECRA Fellowship DE140100598),
  and
  the European Research Council under the
  European Union's Seventh Framework Programme (FP~7)
  \acronym{ERC} Grant Agreement n.~\acronym{[320360, 321035]}).
This research made use of the \acronym{NASA} \project{Astrophysics Data System}.

This project made use of \sdssiii\ data.
Funding for \sdssiii\ has been provided by the Alfred P. Sloan
Foundation, the Participating Institutions, the National Science
Foundation, and the \acronym{U.S.} Department of Energy Office of Science. The
\sdssiii\ web site is http://www.sdss3.org/.

\sdssiii\ is managed by the Astrophysical Research Consortium for the
Participating Institutions of the \sdssiii\ Collaboration including the
University of Arizona, the Brazilian Participation Group, Brookhaven
National Laboratory, Carnegie Mellon University, University of
Florida, the French Participation Group, the German Participation
Group, Harvard University, the Instituto de Astrofisica de Canarias,
the Michigan State/Notre Dame/\acronym{JINA} Participation Group, Johns Hopkins
University, Lawrence Berkeley National Laboratory, Max Planck
Institute for Astrophysics, Max Planck Institute for Extraterrestrial
Physics, New Mexico State University, New York University, Ohio State
University, Pennsylvania State University, University of Portsmouth,
Princeton University, the Spanish Participation Group, University of
Tokyo, University of Utah, Vanderbilt University, University of
Virginia, University of Washington, and Yale University.
	
All of the code written specifically for this project is available in
two open-source code repositories at
\url{https://github.com/andycasey/AnniesLasso} and
\url{https://github.com/davidwhogg/Platypus}.

\software{
	\project{numpy} \citep{numpy},
	\project{scikit-learn}, \citep{sklearn},
	\project{matplotlib} \citep{matplotlib}}.

\end{document}